\documentclass[aps,prb,twocolumn,groupedaddress,graphics,showpacs]{revtex4}
\usepackage{graphics}
\usepackage{epsfig}
\begin{document}
\bibliographystyle{apsrev}

\title{Quantum Shot Noise Suppression in Single-Walled Carbon Nanotubes}
\author{N. Y. Kim}
\thanks{Corresponding author.}
\email[]{Email: nayoung@stanford.edu}
\author{W. D. Oliver}
\altaffiliation{Present address: MIT Lincoln Laboratory,
Lexington, Massachusetts, 02420}
\author{Y. Yamamoto}
\thanks{also at NTT Basic Research Laboratories,
3-1 Morinosato-Wakamiya Atsugi, Kanagawa, 243-01 Japan}
\affiliation{Quantum Entanglement Project, ICORP, JST,\\
E. L.Ginzton Laboratory, Stanford University, Stanford, California
94305}
\author{Jing Kong}
\altaffiliation{Present address:Department of Nano Science and
DIMES, Delft University of Technology, 2628 CJ Delft, The
Netherlands}
\author{Hongjie Dai}
\affiliation{Department of Chemistry, Stanford University,\\
Stanford, California 94305 }
\date{November 19, 2003}

\begin{abstract}
\vspace{0.2in} We study the low frequency current correlations of
an individual single-walled carbon nanotube at liquid He
temperature. We have distinguished two physical regimes -- zero
dimensional quantum dot and one dimensional quantum wire -- in
terms of an energy spacing from the finite tube length in both
differential conductance and shot noise measurements. In a one
dimensional wire regime, we observed a highly suppressed shot
noise from all measured tube devices, suggesting that
electron-electron interactions play an important role.
\end{abstract}


 \maketitle

Single-walled carbon nanotubes (SWNTs) have been an attractive
material for over a decade due to their unique chemical,
mechanical and electronic properties. They are molecular quantum
wires \cite{Dekker99}, an ideal system to probe low-dimensional
physics. Since SWNTs have both spin and orbital degeneracy,
conductance with ideal contacts yield two times of spin-degenerate
quantum unit of conductance, $2(2e^2/h)$. Conductance measurements
of SWNTs coupled to metal electrodes have demonstrated remarkable
electrical transport properties: Coulomb blockade oscillation
\cite{Bockrath97}, the Kondo effect \cite{Nygard00}, ballistic
quantum interference \cite{Liang01,Kong01}, and Luttinger-liquid
behavior \cite{Bockrath99}.

 Noise experiments in carbon nanotubes,
however, are relatively recent for two main reasons: the
difficulty to fabricate nearly-ohmic contacted SWNT devices
\cite{Javey03}, and a technical obstacle to achieve a high
signal-to-noise ratio because of a weak excess-noise signal
embedded in the prevalent background noise. Only a few groups, in
fact, have reported $1/f$ noise of SWNTs and multi-walled
nanotubes \cite{Collins00}, and shot noise measurements on an
ensemble of SWNTs \cite{Rouche02}.

Shot noise refers to the non-equilibrium current fluctuations
resulting from the stochastic transport of quantized charge
carriers. When electron transport is governed by Poisson
statistics as in the random emission of electrons from a reservoir
electrode, the spectral density of the current fluctuations
reaches full shot noise $S_I = 2e\bar{I}$, where $e$ is the
electron charge and $\bar{I}$ is the average current. In a
mesoscopic conductor, quantum shot noise occurs due to the random
partitioning of electrons by a scatterer. If  the noiseless
incoming electrons from a Fermi-degenerate reservoir at zero
temperature are scattered into the outgoing states with a
transmission probability $T$, the transmitted electrons carry
partition noise $S_I = 2e\bar{I}(1-T)$. $T$ is the ratio $T \equiv
\frac{G}{G_Q}$ between the spin-degenerate quantum unit of
conductance $ G_Q =(2e^2/h)\sim (12.5 \text{k}\Omega)^{-1}$ and
the measured DC conductance $G$ for a ballistic conductor
\cite{Buttiker92}. Noise in a quantum system is quantified with
the Fano factor $F \equiv \frac{S_I}{2e\bar{I}}$.

Quantum shot noise has been observed with quantum point contacts
(QPCs) in two-dimensional electron gas systems \cite{Liu98,
Oliver99}. Recently proposed detection schemes for entangled
electrons have uitilized the analysis of shot noise correlations
\cite{Maitre00, Burkard00, Samuelsson03}. Moreover, shot noise
measurements allow one to observe the effective charge $Q$ of
quasiparticles along fractional quantum Hall edge states
\cite{Picciotto97,Comforti02}, yielding  $ S_I = 2Q\bar{I}(1-T)$
and $F \equiv \frac{(1-T)Q}{e}$. Electron-electron interactions in
one-dimensional edge states result in fractionally charged
carriers. This strongly correlated 1D system is modelled as a
``Luttinger liquid" (LL).

A LL exhibits  power-law scaled conductance ( in terms of
temperature and applied voltage), spin-charge separation, and
charge fractionalization \cite{Voit94}.  Quantum Hall edge states
are an example of a chiral LL, in which the forward and
back-scattered electrons move along spatially separate paths. In
contrast, the forward and backward propagating electrons coexist
along a non-chiral LL, such as a SWNT. While noise measurements on
a chiral LL examined directly the effective charge of
quasiparticles, to date there is no noise data for a non-chiral
LL. When electrons travel through a SWNT, an important question is
whether electron-electron interactions in a non-chiral LL are
manifest as a fractional charge. This remains an open question
among  theorists \cite{Bena01,Trauzettel02,Pham02} in the absence
of experimental results. In this Letter, we present the first
experimental results from quantitative, two-port shot noise
measurements of a single isolated SWNT at 4 K, which may help to
clarify this issue.

\begin{figure}
\epsfig{figure=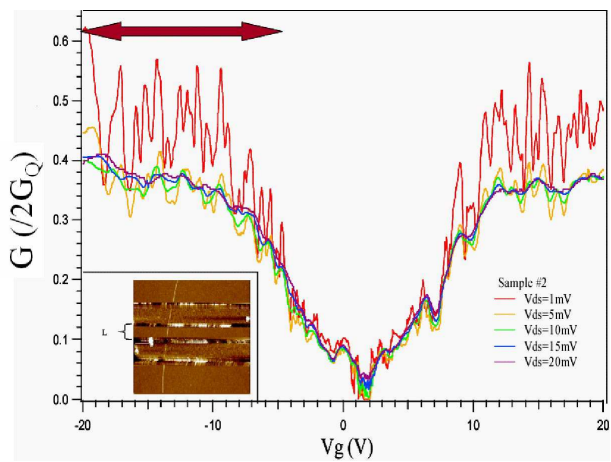,width = 3.2 in} \caption{(Color online)
Conductance ($G = I/V_{ds}$) vs the gate voltage ($V_g$) with the
different drain-source voltages ($V_{ds}$) at 4 K . Conductance is
normalized by $2G_Q = 2(2e^2/h)$. Shot noise measurements were
performed by sweeping $V_g$ where the conductance is constant,
indicated in the red arrow region. (Inset)Atomic force microscopy
image of the SWNT device. \label{Fig:fig1}}
\end{figure}

An  SWNT was synthesized as reported previously \cite{Soh99}: An
individual SWNT was grown by Alumina-based chemical vapor
deposition method onto a 500 $\mu $m -thick SiO$_2$ layer which
was thermally grown on a heavily doped Si substrate. The Si
substrate was used as the back gate. The metal electrodes were
patterned by electron beam lithography, defining a device length
between 200 and 600 nm (Inset of Fig.~\ref{Fig:fig1}). Ti/Au or
Ti-only metal electrodes featured good contacts for low resistance
SWNT devices. Atomic force microscopy imaging enabled us to select
the device consisting of a single isolated SWNT which was 1-3 nm
diameter and 200 - 600 nm long.  We first measured conductance as
a function of gate voltage ($V_g$) to classify the tube as
metallic or semi-conducting. The metallic tubes were chosen for
noise measurements. Resistances of the selected devices were
typically 15 $\sim$ 50 k$\Omega$ at room temperature and about
12.2 $\sim$ 25 k$\Omega$ at 4 K.

\begin{figure}
\epsfig{figure=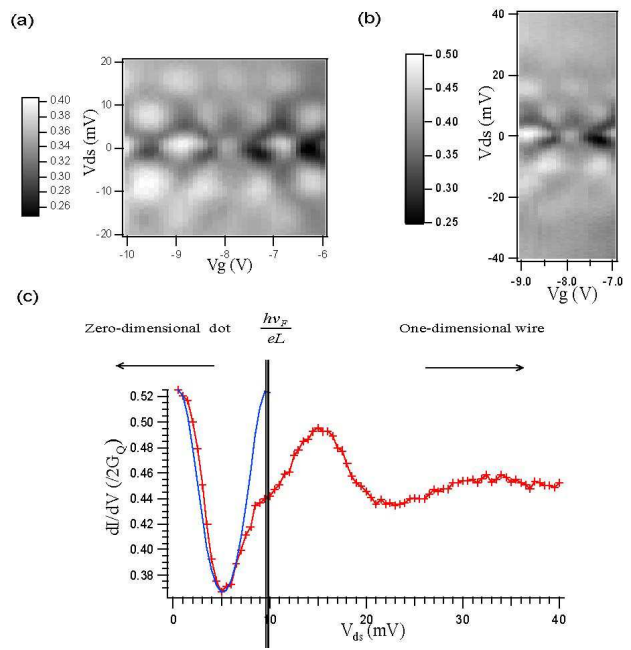,width = 3.2 in}
\caption{(Color online) (a)(b) The differential conductance
$dI/dV_{ds}$ as a function of the drain-source voltage ($V_{ds}$)
and the gate voltage ($V_g$). (c) Theoretical g from the simple
double-barrier Fabry-Perot cavity model (blue) and the
experimental result (red cross) at $V_g =-5$V. \label{Fig:fig2}}
\end{figure}

Predictions \cite{White98,McEuen99} and experiments
\cite{Bockrath97,Liang01,Kong01,Tans98} have shown that both the
elastic and the inelastic mean free path are at least on the order
of microns in metallic nanotubes at low temperatures. Electron
transport within $200 - 600$ nm-long SWNTs is believed to be
ballistic \cite{Liang01,Kong01}. We have observed a quantum
interference pattern in the differential conductance $dI/dV_{ds}$
(where $V_{ds}$ is the drain-source voltage) as a function of
$V_{ds}$ and $V_g$ at 4K as shown in Fig.~\ref{Fig:fig2}(a). The
diamond structures are determined by the finite length of the tube
between the metal electrodes. The sample is 360 nm-long, and the
corresponding energy spacing is roughly 10 meV. The size of the
pattern is consistent with the energy spacing.  The interference
structure continues until approximately $V_{ds} \sim 20$ mV, above
which it fades away (Fig.~\ref{Fig:fig2}(b)). This interference
pattern can be explained by a resonant tunnelling model with
quantized energy levels due to finite reflection at the
electrode-SWNT interfaces (Fabry-Perot cavity effect)
\cite{Liang01}.

We consider a simple model to understand the behavior of
$dI/dV_{ds}$. Suppose the interfaces act as potential barriers
with constant transmission probability $T_L$ and $T_R$ .  The
energy dependent overall transmission probability $T(E)$ is
written for a double-barrier structure as\\
$$T(E)=
\frac{T_LT_R}{1+(1-T_L)(1-T_R)-2\sqrt{(1-T_L)(1-T_R)} \cos
\Theta(E)},$$

where $\Theta(E)$ is the phase accumulation caused by multiple
reflections and  a function of $V_{ds}$ and $V_g$. This simple
model fits well the experimental $dI/dV_{ds}$ data for a small
$V_{ds}$ energy, but deviates significantly as $V_{ds}$ increases
(Fig.~\ref{Fig:fig2}(c)). We can interpret this conductance
behavior in terms of the tube length $L$ and $V_{ds}$. The quantum
state energy spacing ($\Delta E$ ) from the longitudinal
confinement is inversely proportional to $L$, $\Delta E \sim
\frac{hv_F}{L}$, where $v_F = 8 \times 10^5$ m/s is the Fermi
velocity. When $V_{ds}$ is smaller than $\frac{\Delta E}{e}$, the
coherence length of electron wavepackets is longer than L. In this
region, the SWNT operates as an isolated zero-dimensional quantum
dot between two leads. For $V_{ds} > \frac{\Delta E}{e}$, the
coherence length of electron wavepackets is shorter than the tube
length. In this limit, each wavepacket passes through the
one-dimensional conductor, and the oscillating period apparently
increases. This phenomenon may indicate the correlated electrons
\cite{Peca03}.

\begin{figure}
\epsfig{figure=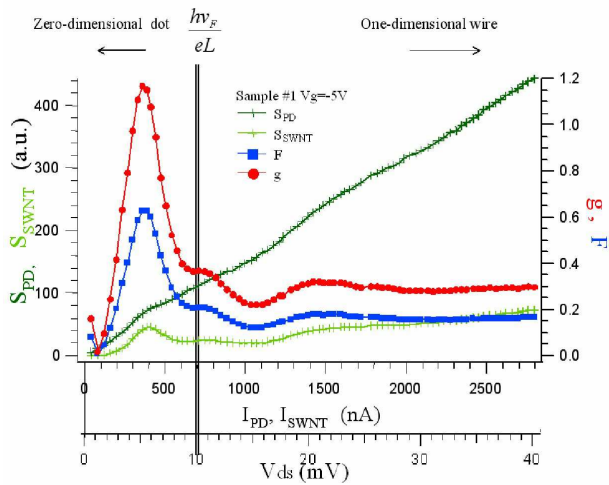,width=3.2in} \caption{(Color online)
Representative data of the LED/PD noise ($S_{PD}$, dark green),
the SWNT noise ($S_{SWNT}$, light green), Fano factor (F, blue)
and shot noise suppression factor (g, red). \label{Fig:fig3}}
\end{figure}

The  two-terminal shot noise measurements were implemented by
placing two current noise sources in parallel: one is  a weakly
coupled light emitting diode (LED) and photodetector (PD) pair,
and the other is a SWNT. The LED/PD pair, whose coupling
efficiency from the LED input current to the PD output current was
~ 0.1 $\%$ at 4 K, served as the full shot noise source. The high
signal-to-noise-ratio was achieved by implementing an AC
modulation lock-in technique and a resonant circuit together with
a home-built cryogenic low-noise preamplifier
\cite{Liu98,Oliver99}. The input-referred voltage noise was
approximately $2.2$ nV$/\sqrt{\text{Hz}}$ at 4 K with a resonance
frequency  $\sim 15$ MHz. The signal was fed into a $9$ MHz
bandpass filter followed by a square-law detector and a lock-in
amplifier. The Fano factor $F(I_0) \equiv \frac{S_{SWNT}}{S_{PD}}$
was obtained from the ratio of the tube noise ($S_{SWNT}$) and the
full shot noise ($S_{PD}$) at various currents $I_0$. The current
noise generated in the LED/PD pair was measured while the SWNT was
voltage-biased at an un-modulated DC current $I_0$ . Figure
~\ref{Fig:fig3} presents typical data of the measured LED/PD and
SWNT shot noise and Fano factor. Two distinct regimes are
noticeable depending on the bias voltage $V_{ds}$. At a low bias
voltage ($0 < V_{ds} < 10$ mV), $F$ and $g$ oscillates as a
function of $V_{ds}$. For a high bias voltage ($V_{ds}> 10 $mV),
$F$ and $g$ reach saturated values.

The SWNT current fluctuations show many features and can be
divided into two distinct regimes according to $\Delta E$.  A
non-interacting electron model predicts that an expected Fano
factor is $1-T(I_0)$ at a net transmission probability $T(I_0)$.
This simple picture, however, does not match the measured $F(I_0)$
over the entire $V_{ds}$ range. In a low $V_{ds}$, $F(I_0)$
deviates from the non-interacting prediction in a complicated way;
the suppressed shot noise around zero $V_{ds}$ becomes enhanced.
Enhanced noise would be a consequence of correlated resonant
tunnelling \cite{Safonov03}. Currents through coherent resonant
states may be affected by weak localized states within the system
such that additional noise occurs due to interaction. While
conductance does not reflect this effect, the shot noise manifests
it sensitively.

The SWNT noise in a one-dimensional wire regime is suppressed far
below the non-interacting  electron prediction.  For instance, the
tube in Fig.~\ref{Fig:fig3} has $R \sim$ 14.3k$\Omega$ at $V_{ds}
=$ 40 mV corresponding to $T \sim 0.45$ and $1 - T \sim 0.55$;
however, the observed $F$ value at $V_{ds}= 40$ mV is 0.16.
Defining this shot noise suppression $g = F/(1-T)$, the SWNT noise
is then written by $S_I = 2e\bar{I}g(1-T)$. The suppression factor
$g$ (red in Fig.~\ref{Fig:fig3}.) falls between 0.2 and 0.3 for a
large $V_{ds}$, which matches the theoretical value of the
Luttinger parameter in SWNTs \cite{Egger97, Kane97}.

\begin{figure}
\epsfig{figure=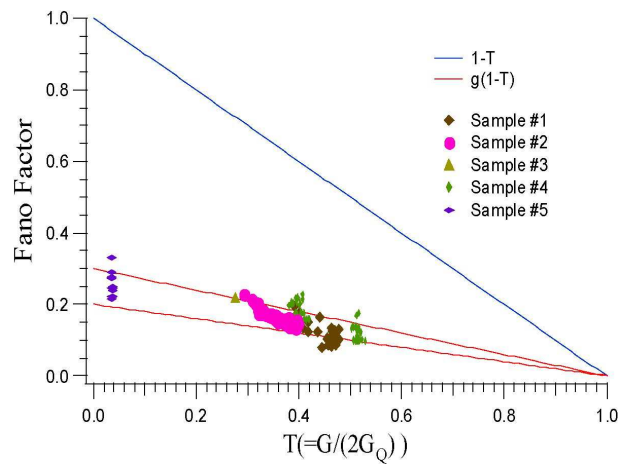,width=3.2in} \caption{ (Color online) Fano
factor ($F$) of five different metallic devices as a function of
the transmission probability $T$. We plot the noninteracting  $F
(= 1-T)$ (dark blue) and the interacting models $F = 0.2(1-T)$
(red) and $F = 0.3(1-T)$ (red) with samples 1,2,3,4 and 5.
\label{Fig:fig4}}
\end{figure}

Figure ~\ref{Fig:fig4} summarizes the Fano factors of five
different metallic SWNTs (samples 1-5) as a function of the
transmission probability $T$. All values are extracted from the
one-dimensional wire regime. For each device, the shot noise
measurements were performed by varying $V_g$. All five tubes
manifested both a strong suppression and a consistent $g$ value.
Surprisingly, even a highly resistive tube also showed a similar
suppression (sample 5).

Two different models are considered in an attempt to explain the
results. First is the coherent scattering theory
(Landauer-Büttiker formalism) \cite{Buttiker92} of the
non-interacting electrons. $F$ in this picture (the blue solid
line in Fig.~\ref{Fig:fig4}) clearly mismatches the experimental
result. As the other model, we regard the suppressed shot noise in
the context of a Luttinger-liquid model. Although an exact model
for the current noise in a non-chiral LL is absent, we
phenomenologically combine the partition noise expected from the
non-interacting picture with a Luttinger parameter. The Luttinger
parameter is predicted to have a value $g \in$ [0.2, 0.3] in
carbon nanotubes \cite{Egger97,Kane97} and, indeed, it was
experimentally determined to be $0.26 - 0.28$ from a power-law fit
of the conductance in tubes weakly coupled to Fermi reservoirs
\cite{Bockrath99}. We plotted two linear lines by identification
of $Q = ge$, i.e. $F = 0.2 (1-T)$ and $F = 0.3 (1-T)$ based on the
limiting g values in Fig.~\ref{Fig:fig4}. Most data points fall
between the two lines. In order to test the universality of the
additional shot noise reduction factor, the further studies should
be needed, for example, the effect of the metal electrode
material, the interface between electrodes and tubes and the
growth condition of SWNTs.

We would like to point out that $F$ of SWNTs increases as $T$
decreases in a similar manner to the ballistic conductor except
with the additional suppression, which might be related to the
electron-electron interactions within the tube. All Fano factors
in Fig.~\ref{Fig:fig4} were taken at the high $V_{ds} \gg
\frac{hv_F}{eL}$ . The non-equilibrium noise in an interacting
system has not been studied theoretically in this high-energy
excitation regime to our best  knowledge. A rigorous theoretical
work is now in demand to explain the shot noise in individual
SWNTs.

We acknowledge Prof. Quate for his encouragement and atomic force
microscopes for imaging the SWNT devices and thank the ARO-MURI
grant DAAD19-99-1-0215 for support.

\end{document}